\documentclass[journal]{IEEEtran}
\usepackage{graphicx}
\usepackage{cite}
\usepackage{picinpar}
\usepackage{amsmath}
\usepackage{url}
\usepackage[latin1]{inputenc}
\usepackage{colortbl}
\usepackage{soul}
\usepackage{multirow}
\usepackage{pifont}
\usepackage{color}
\usepackage{alltt}
\usepackage[hidelinks]{hyperref}
\usepackage{enumerate}
\usepackage{siunitx}
\usepackage{breakurl}
\usepackage{epstopdf}
\usepackage{pbox}
\usepackage[table,xcdraw]{xcolor}
\usepackage[]{algorithm2e}
\usepackage{comment}

\ifCLASSINFOpdf
\else
\fi

\begin{document}
%
\title{A Wearable CMOS Biosensor with 3 Designs of Energy-Resolution Scalable Time-Based Resistance to Digital Converter}
%
%

\author{Dong-Hyun~Seo,
        Baibhab~Chatterjee,~\IEEEmembership{Student Member,~IEEE,}
        Sean Scott,
        Daniel Valentino,~\IEEEmembership{Member,~IEEE,}
        Dimitrios Peroulis,~\IEEEmembership{Fellow,~IEEE,}
        and~Shreyas~Sen,~\IEEEmembership{Senior~Member,~IEEE}
        
}


\maketitle

\begin{abstract}
This paper presents the design and analysis of a wearable CMOS biosensor with three different designs of energy-resolution scalable time-based resistance to digital converters (RDC), targeted towards either minimizing the energy/conversion step or maximizing bit-resolution. The implemented RDCs consist of a 3-stage differential ring oscillator which is current starved with the resistive sensor, a differential to single ended amplifier, an off-chip counter and serial interface. The first design RDC included the basic structure of time-based RDC and targeted low energy/conversion step. The second design RDC aimed to improve the rms jitter/phase noise of the oscillator with help of speed-up latches, to achieve higher bit-resolution as compared to the first design RDC. The third design RDC reduced the power consumption by scaling the technology with the improved phase-noise design, achieving 1-bit better resolution as that of the second design RDC. Using a time-based implementation, the RDCs exhibit energy-resolution scalablity, and consume 861nW with 18-bit resolution in design 1 in TSMC 0.35$\mu$m technology. Design 2 and 3 consume 19.1$\mu$W with 20-bit resolution using TSMC 0.35$\mu$m, and 17.6$\mu$W with 20-bit resolutions using TSMC 0.18$\mu$m, respectively (both with 10ms read-time, repeated every second). With 30ms read-time, design 3 achieves 21-bit resolution, which is the highest resolution reported for a time-based ADC. The 0.35$\mu$m time-based RDC is the lowest-power time-based ADC reported, while the 0.18$\mu$m time-based RDC with speed-up latch offers the highest resolution. The active chip-area for all 3-designs are less than 1.1 mm\(^{2}\).
\end{abstract}

\IEEEpeerreviewmaketitle
\begin{figure}[!t]
\centering
\includegraphics[width=3.5in]{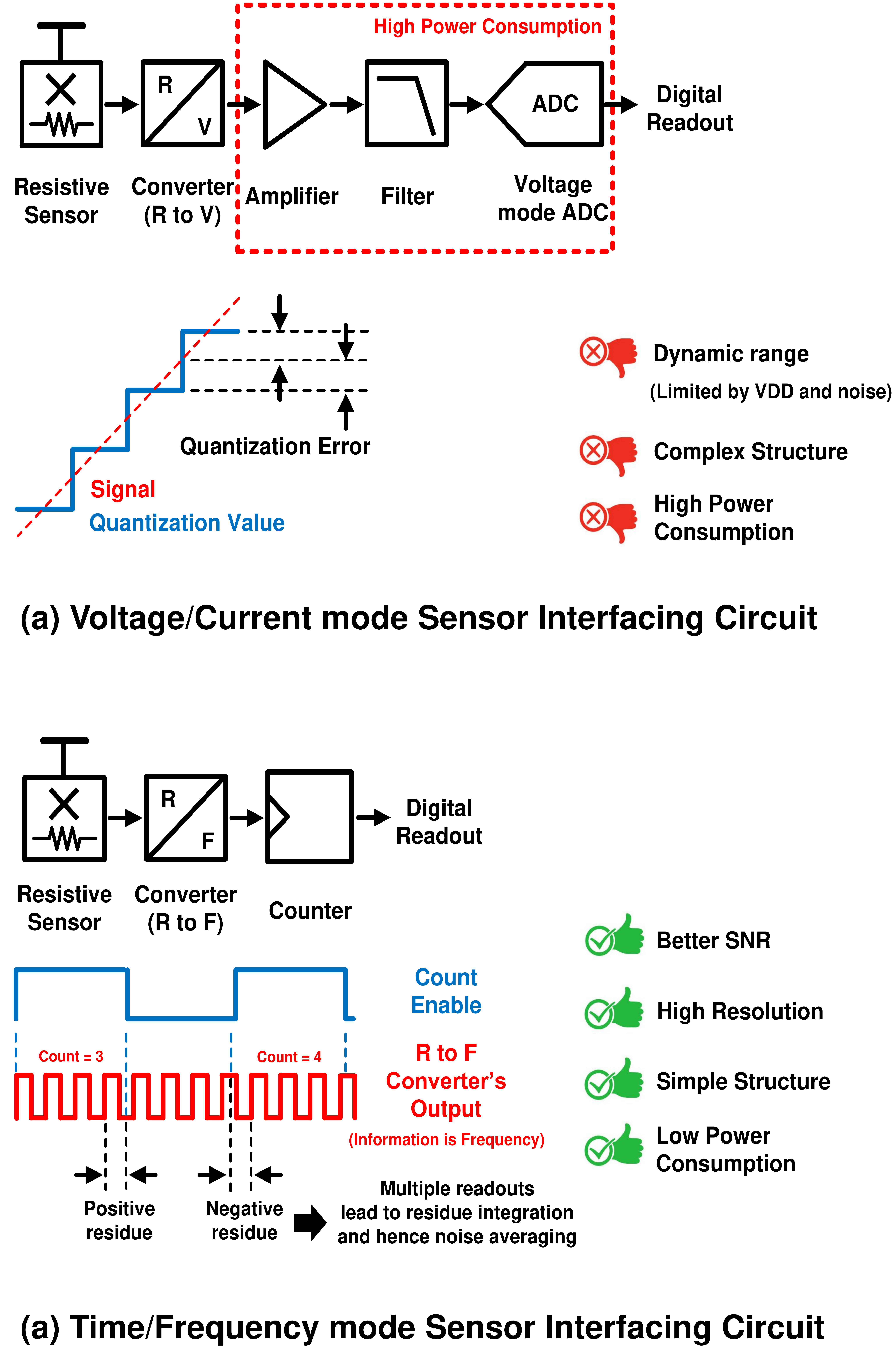}
\caption{Inherent advantages provided by time/frequency mode ADCs over voltage/current mode ADCs for low-speed high-resolution resistive sensing applications: (a) traditional voltage/current mode ADC sensor interfacing circuits, (b) time/frequency mode ADC sensor interfacing circuits.}
\label{VM_vs_TM_Compare}
\end{figure}
\begin{figure*}[!t]
\centering
\includegraphics[width=7in]{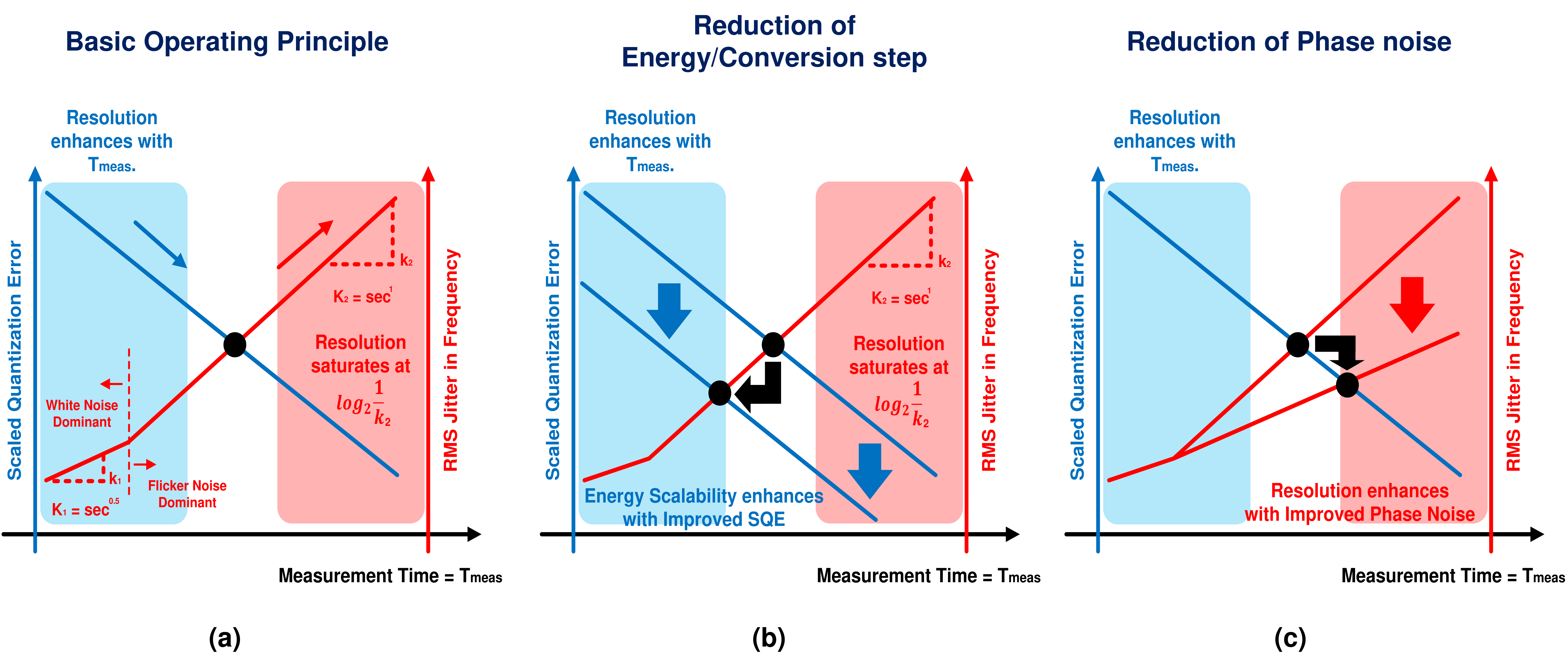}
\caption{Operating principle of time-based RDC: (a) resolution trade-offs between scaled quantization noise and jitter/phase noise of the ring oscillator in the design: quantization error decreases with time, whereas jitter accumulates with time, resulting in a saturated maximum achievable resolution in the jitter dominated region (b) reduction in measurement time (and hence measurement energy and energy/conversion step) by reducing the scaled quantization error of the ring oscillator, (c) improvement in resolution resulting from improving jitter/phase noise of the ring oscillator}
\label{operating_principle_of_RDC}
\end{figure*}
\section{Introduction}
\label{intro}
\IEEEPARstart{I}{n} recent years, low-power wearable bio sensing devices have presented great potentials for many biomedical applications including diagnosis, physiological monitoring and health care systems \cite{biosensor1}-\cite{biosensor5}. The pursuit of convenience for continuous monitoring in both hospital and ambulatory applications requires these devices to have a small form of factor and low power consumption for battery-powered wearable/portable use. Thus, integrated circuits (ICs) are needed in biomedical fields in order to create wearable medical sensing applications. The most critical requirement of medical devices is the accurate transmission of data/information, which requires high bit-resolution. The key challenge in designing high resolution biomedical ICs derives from system/circuit/ambient noise and power, exhibiting trade-offs among achievable resolution, noise and power \cite{trade-off6}.

Fig. 1 shows the fundamental advantages provided by time/frequency mode analog-to-digital converter (ADC) over voltage/current mode ADC for low speed and high resolution applications in the noise and supply voltage limited regime. Although digital signal processors and integrated circuits take advantage of technology scaling to achieve improvements in power, speed, size, and cost, scaling of supply voltage cause a significant disadvantage to the available voltage dynamic range. In addition, the voltage/current mode ADC interface requires signal conditioning circuits such as analog amplifiers and filters between the sensor devices and the ADC, making it challenging to reduce power. Also, voltage/current mode ADCs need sophisticated noise cancelling techniques to diminish the quantization noise as displayed in Fig. 1(a), which improves the signal to noise ratio (SNR) and bit-resolution. On the other hand, the reduction of gate delays has led an improvement in "time-resolution" in scaled devices. Furthermore, the time/frequency mode ADC can achieve high resolution with increased enable time which can integrate residues over more time \cite{CICC7}-\cite{JSSC7}, leading to a time-domain averaging of the noise which is displayed in Fig. 1(b). Thus, sensing the data/information through time-based techniques (which is a time difference between two rising or falling edges) can potentially represent a better solution than sensing in voltage mode (which is a difference between two node voltages), when ADCs are implemented in a scaled process \cite{time-base ADC8}. The time domain ADC can be as simple as a ring oscillator (that converts resistance to frequency by starving the ring oscillator with the resistive sensor), the output frequency of which can be provided to a multi-bit digital counter with a predefined enable time. The output of the counter would be a direct digital representation of the resistance.

Along With the popularization of biomedical applications and their increasing demands, devices with low power consumption and high resolution sensor have been increasingly preferred. Resistive sensors possess numerous strengths, including good stability, low cost, and ease to be interfaced by readout circuits. Due to these strengths, resistive sensors have been extensively practiced and utilized in diverse fields such as physiological monitoring, environmental and biomedical analysis \cite{resistive sensor9}-\cite{resistive sensor12}. As devices that include a resistive sensor are widely adopted in biomedical applications with diverse dynamic range requirements (such as temperature, pressure, and radiation), this paper aims to analyze energy-resolution scalability of the proposed time-based resistance-to-digital converter (RDC).

In this paper, in order to resolve the challenges and address the aforementioned advantages, the CMOS biosensor is implemented as a resistive sensor and time-based RDC. This paper is organized as follows. Section II describes the operating principle and system architecture of three designs of energy-resolution scalable time-based RDC. Section III presents the details of circuit design with the simulation. Section IV explains and describes the system level simulation and result of time-based RDC. Section V shows experimental results of the implemented chips along with a comparison with the state-of-the-art. Finally, concluding remarks are presented in Section VI.
\section{System Architecture}
\label{theory}
\begin{figure}[!t]
\centering
\includegraphics[width=3.5in]{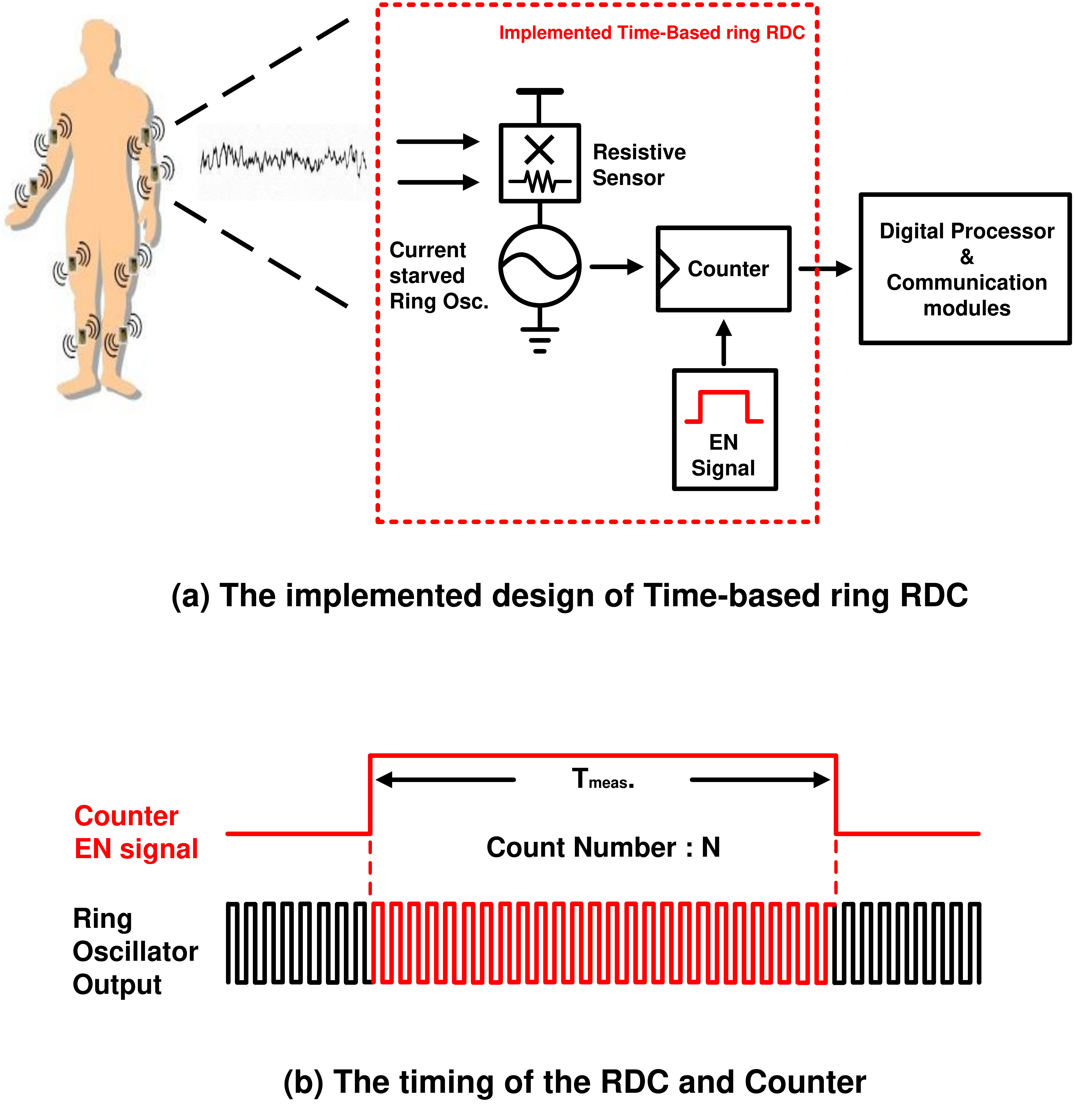}
\caption{System architecture: (a) Simplified diagram of time based Ring-RDC based biosensor node, (b) the timing of the RDC and counter}
\label{Structure}
\end{figure}
\begin{figure*}[!t]
\centering
\includegraphics[width=7in]{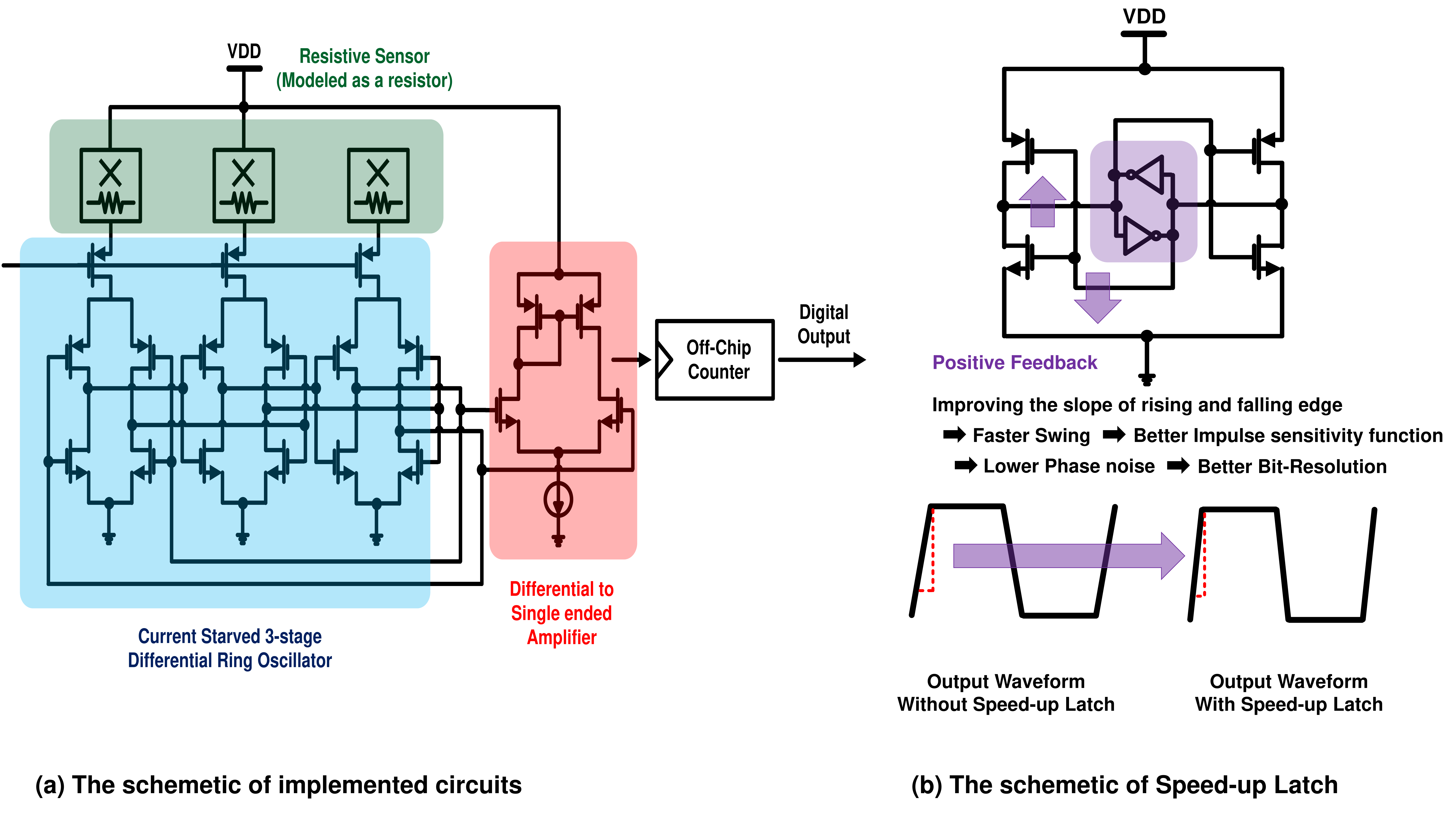}
\caption{Implemented circuit schemetic: (a) a resistive sensor, a current starved 3-stage differential ring oscillator and a differential to singled ended amplifier and (b) improving the slope of rising and falling edge of a 3-stage ring oscillator using speed-up latch.}
\label{circuit_schemetic}
\end{figure*}

Fig. 2 shows the operating principle of CMOS biosensor with time-based architectures that help to achieve better bit-resolution with more measurement time \((T_{meas.})\). The bit-resolution of the time-based RDC is determined by the scaled quantization error (which is the ratio of one counting cycle with the total measurement time, i.e. \(\frac{T}{T_{meas.}}\)) and the accumulated jitter around desired frequency measured as described in Fig. 2(a). The scaled quantization error (SQE) decreases with an increase in the \((T_{meas.})\), where T refers to a time-period of oscillator. On the contrary, the accumulated rms jitter/phase noise of the ring oscillator linearly increases with \(T_{meas.}\) When the overall phase noise is dominated by the flicker noise of the tail current sources which slowly change, the cycles of oscillation of oscillator constantly change, either accelerating or decelerating, due to the correlated supply and substrate noise. It ends up accumulating a large difference in phase. The correlated nature of noise generates the increased difference quadratically with the total time of accumulation. This results in a rms jitter proportional to the \(T_{meas.}\)\cite{jitter/phasenoise13}-\cite{jitter/phasenoise14}. Combining the effects of SQE with jitter/phase noise, we can write the total scaled quantization error with jitter (SQEJ) as given by Eq. 1. 
\begin{equation}
\begin{aligned}
	SQEJ = \frac{T+k \times T_{meas.}}{T_{meas.}}
\end{aligned}
\label{comm_eqn}
\end{equation}
where k refers to the slope of the linearly accumulating rms jitter/phase noise with \(T_{meas.}\) The bit-resolution can be defined by Eq.2.
\begin{equation}
\begin{aligned}
	bit\textunderscore resolution = log_{2}(\frac{1}{SQEJ})
\end{aligned}
\label{comm_eqn}
\end{equation}
Even though \(T_{meas.}\) increases, the bit-resolution is eventually saturated at \(log_{2}(\frac{1}{k}) \). There are two ways to enhance the energy-resolution trade-offs in this architecture. One way is to improve the absolute value of the SQE, as shown in Fig 2(b), which results in an unchanged maximum achievable bit-resolution at a lower \((T_{meas.})\), thereby reducing the energy required for measurement as well as the energy per conversion step. The other way is to improve the rms jitter/phase noise as shown in Fig 2(c), which will result in a lower slope and better bit-resolution. In this paper, we focus on the latter way of improving rms jitter/phase noise for increasing the bit-resolution from design 1 to design 2 and 3 of the proposed architecture. The first design provided the basic structure of time-based RDC which aimed at having a low energy consumption \cite{CICC7}-\cite{JSSC7}. In the second design, the improved rms jitter/phase noise technique was implemented with the help of speed-up latches. The third design also used the improved rms jitter/phase noise technique in a scaled technology that offered lower power consumption.

Fig. 3 shows the system architecture. The simplified system architecture of the wearable node, composed of a resistive sensor, a current starved ring oscillator and a digital counter is described in Fig. 3(a). The counter is implemented off-chip. The current starved ring oscillator converts sensor resistance value to oscillation frequency of current starved ring oscillator as a clock output. The clock is supplied to the counter. Fig. 3(b) describes the timing of the RDC and counter. During measurement time (\(T_{meas.}\)), the counter counts the rising edges of output. The counter output represents the integer number of output cycles in one readout period.
\section{Circuit Design}
\subsection{0.35$\mu$m Time-Based RDC}
Fig. 4(a) shows the 0.35$\mu$m time-based RDC which includes resistive sensor devices which are integrated in the same chip, a current starved 3-stage differential ring oscillator (DRO) and a differential to single ended amplifier in order to convert the resistance to frequency, which directly depends on the delay introduced by each inverter stage. Limiting the amount of current is a way to control the delay. In this architecture, the resistive sensors are designed in a way that the generated frequency primarily depends on the amount of current allowed by the resistance, and not on other factors such as the load capacitance \(C_{L}\). The concepts of Impulse Sensitivity Function (ISF) and Noise Modulating Function (NMF) are critical to understand that the phase noise at a particular offset increases with the number of stages for a DRO with given power dissipation and frequency. Nonetheless, the phase noise for the single ended ring oscillator SRO works in a different way. Different from the phase noise for the DRO, it does not increase with the number of stages. \cite{jitter/phasenoise13}-\cite{jitter/phasenoise14}. In the case of a SRO, the amount of phase noise does not correlate with the number of stages. However, a SRO has an issue in that it is influenced by common mode supply and substrate noise too easily. Nevertheless, a DRO is more robust against common mode supply and substrate noise even in situations where the phase noise increases linearly with number of stages. Due to this feature, a DRO was chosen over a SRO in this paper, and in order to minimize the issues a DRO has with regard to the phase noise, the number of stage was fixed at a minimal required number (3) for implementation of a DRO. Symmetry was considered to be an important factor during layout as it contributes to minimizing the effects of supply and substrate noise. The differential to single ended converter which is connected to the output of the 3-stage DRO does not require voltage gain, since the input of the differential to single ended converter is rail-to-rail. This means that the transconductance ($g_m$) requirement of the differential to single ended converter is small. As a result, the power consumption of the differential to single ended converter is small with extremely relaxed design constraints, which makes the overall design almost digital and scaling friendly. The 0.35um time-based RDC has the phase noise of -92.5dBc/Hz @ 100kHz offset and oscillation frequency of 83.2 MHz with 94.4uW power consumption (when continuously on) in simulation\cite{CICC7}-\cite{JSSC7}.

\subsection{0.35$\mu$m Time-Based RDC with Speed-up Latch}
Fig. 4(b) shows how speed-up latch is implemented with the 0.35$\mu$m time-based RDC. The single-side band (SSB) phase noise of the DRO is defined by Eq.3-5 [12].
\begin{equation}
\begin{aligned}
	L(\Delta f) = \frac{8}{3\eta} N \frac{kT}{P} (\frac{V_{DD}}{V_{Char}}+\frac{V_{DD}}{R_{L}I_{tail}}) \frac{f_{o}^{2}}{\Delta f^{2}}
\end{aligned}
\label{comm_eqn}
\end{equation}
\begin{equation}
\begin{aligned}
	V_{Char\textunderscore long\textunderscore channel} = \frac{\Delta V}{\gamma}
\end{aligned}
\label{comm_eqn}
\end{equation}
\begin{equation}
\begin{aligned}
	V_{Char\textunderscore short\textunderscore channel} = \frac{E_{c}L}{\gamma}
\end{aligned}
\label{comm_eqn}
\end{equation}
where \(\eta\) is the ratio of stage delay of rising/falling time, N is the number of DRO stages, k is the Boltzmann constant, T is temperature, P is the power dissipation, \(R_{L}I_{tail}\) is the output swing, \(f_{o}\) is the output frequency, and \(\Delta f\) is the offset frequency at which phase noise is calculated. Increasing power reduces phase noise, and reducing frequency of operation will also reduce phase noise at a particular offset. Improving the slope of rising and falling edge of a 3-stage ring oscillator enhances the phase noise performance by improving the swing. Using speed-up latches at the output of each stage of the DRO, we improve the white-noise induced phase noise to -124.5 dBc/Hz @ 1MHz offset, as verified through simulations. The speed-up latch provides positive feedback, improving the slope of rising and falling edge of a 3-stage ring oscillator as shown in Fig. 4(b). In Fig. 5(a), (b) and (c), the simulated power consumption, frequency and phase noise are presented, respectively, as a function of the sizing of the transistors in the speed-up latch and the DRO. Since the current is limited by the resistive sensor, increasing the size of the DRO transistors capacitively loads the circuit and reduces the frequency. However, since the output of the latch provides positive feedback, the slope of output of each stage of the DRO increases (increasing the swing and frequency) which is described in Fig 5(b). The phase noise improves with the level of power dissipation. However, the best design points are different in terms of power consumption, frequency and phase noise in 0.35$\mu$m time-Based RDC with Speed-up Latch. For this reason, a cost function is defined by Eq.6.
\begin{figure}[!t]
\centering
\includegraphics[width=3.5in]{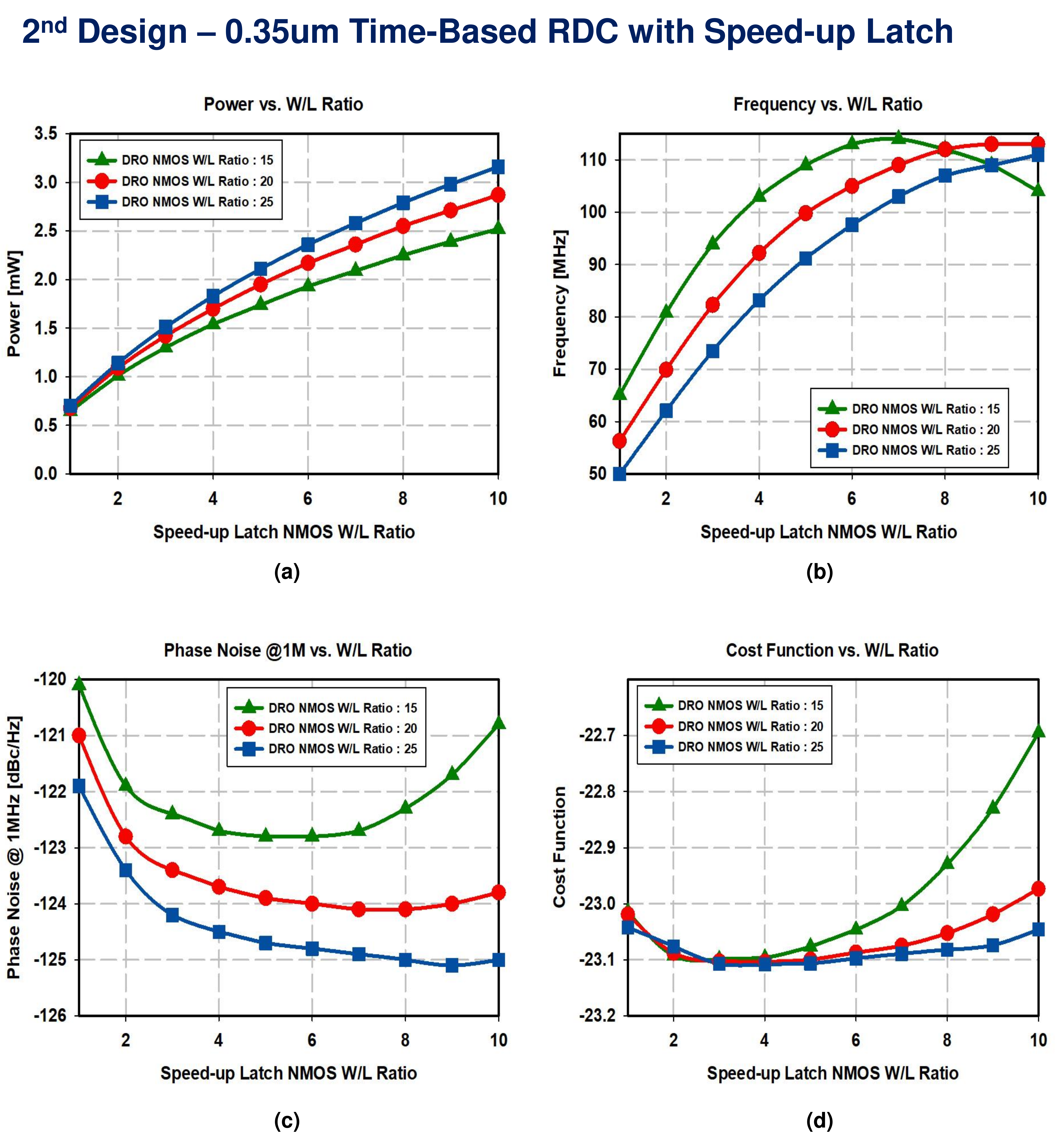}
\caption{Simulation results for the time-based RDC in 0.35$\mu$m technology: (a) power, (b) frequency, (c)phase noise and (d) minimization cost function vs. sizing of the DRO transistors.}
\label{Sim_350nm}
\end{figure}
\begin{equation}
\begin{aligned}
	Cost\textunderscore Function = log_{10}(\frac{10^{PN/10}}{F}\times P)
\end{aligned}
\label{comm_eqn}
\end{equation}
where PN, F and P represent the scaled linear phase noise, frequency and power of the DRO, respectively. Reducing this cost function would mean reducing PN at a high frequency of operation, but at lower power. In Fig. 5(d), the minimized cost function and the optimized sizing of the speed-up latch and sizing of the W/L ratio of the NMOS in the ring oscillator is presented. For the best performance, the NMOS (W/L) ratios of the speed-up latch is determined to be 4 and the NMOS (W/L) ratios of the DRO is determined to be 25 at 83.2 MHz with 1.83 mW power consumption (when continuously on) in simulation.
\begin{figure}[!t]
\centering
\includegraphics[width=3.5in]{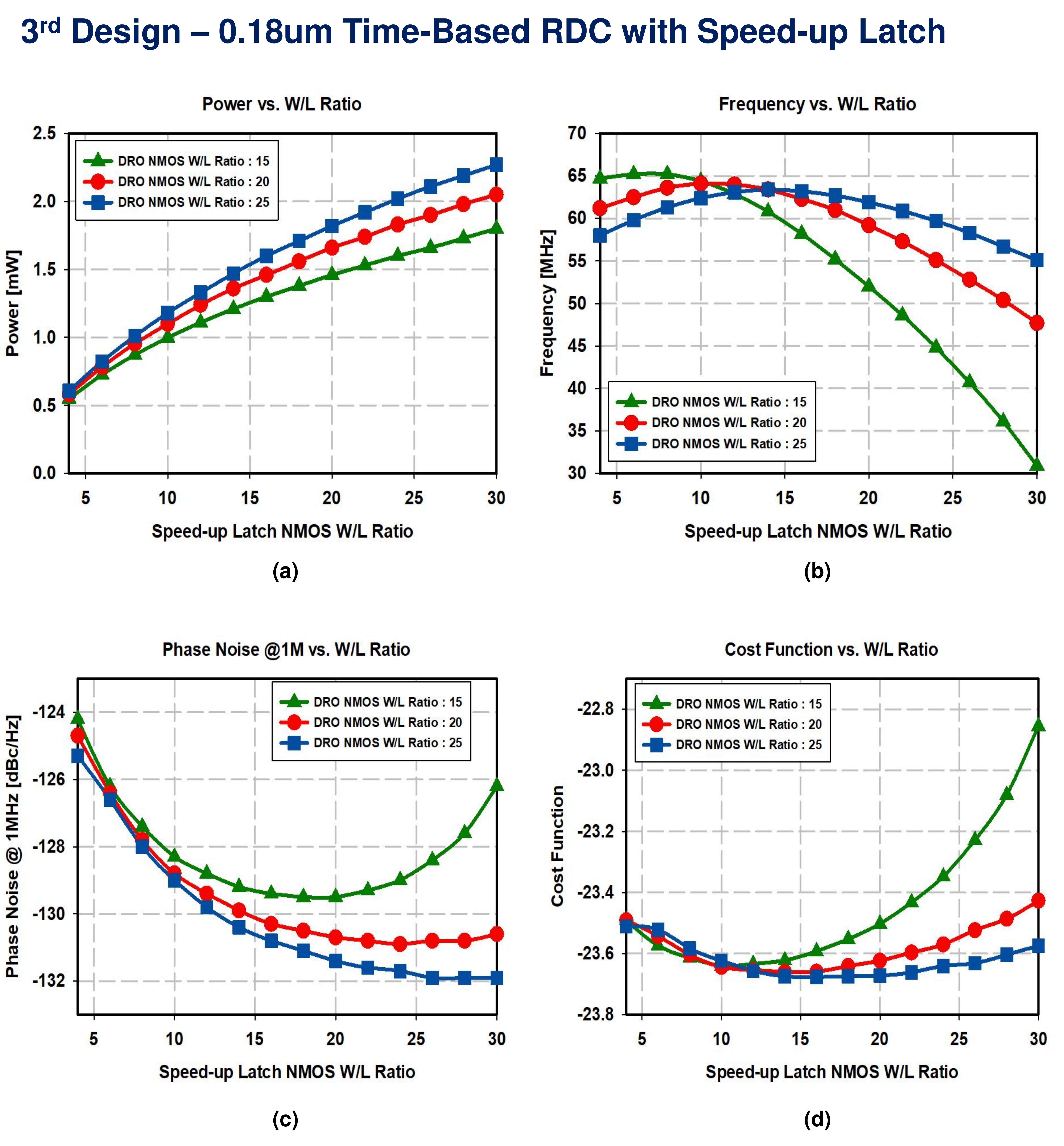}
\caption{Simulation results for the time-based RDC in 0.18$\mu$m technology: (a) power, (b) frequency, (c) phase noise and (d) minimization cost function vs. sizing of the DRO transistors.}
\label{Sim_180nm}
\end{figure}
\subsection{0.18$\mu$m Time-Based RDC with speed-up latch}
Technology scaling is the decisive factor that lead to high performance circuit. Threshold voltage of the device must be reduced proportionally as supply voltage reduces to sustain the output performance of transistor. Technology scaling has also reduced the gate delay, the parasitic capacitances, and the energy and active power per transition. From the Eq.(3), Improving the delay of a 3-stage ring oscillator enhances the phase noise performance by improving the ratio of stage delay of rising/falling time.  Scaling down from 0.35$\mu$m to 0.18$\mu$m technology with the use of speed-up latches at the output of each stage of the DRO improves the power dissipation and the phase noise to 1.3mW (when continuously on) and -130.5 dBc/Hz @ 1MHz offset, respectively, as verified through simulations. In Fig. 6(a), (b) and (c), the simulated power consumption, frequency and phase noise are presented. More specifically, Fig. 6(a), (b), and (c) show how the simulated power consumption, frequency, and phase noise change according to the size of the transistor in the speed-up latch and the DRO in 0.18um. Compared to 0.35$\mu$m time-based DRO with speed-up latch, power dissipation and phase noise are improved, which accordingly improved the cost function. In Fig. 6(d), the minimized cost function and the optimized sizing of the speed-up latch and sizing of the W/L ratio of the NMOS in the ring oscillator is presented. For the best performance, the NMOS (W/L) ratios of the speed-up latch is determined to be 16 and the NMOS (W/L) ratios of the DRO is determined to be 25 at 61.3 MHz with 1.3 mW (when continuously on) power consumption in simulation.
\begin{figure}[!t]
\centering
\includegraphics[width=3.5in]{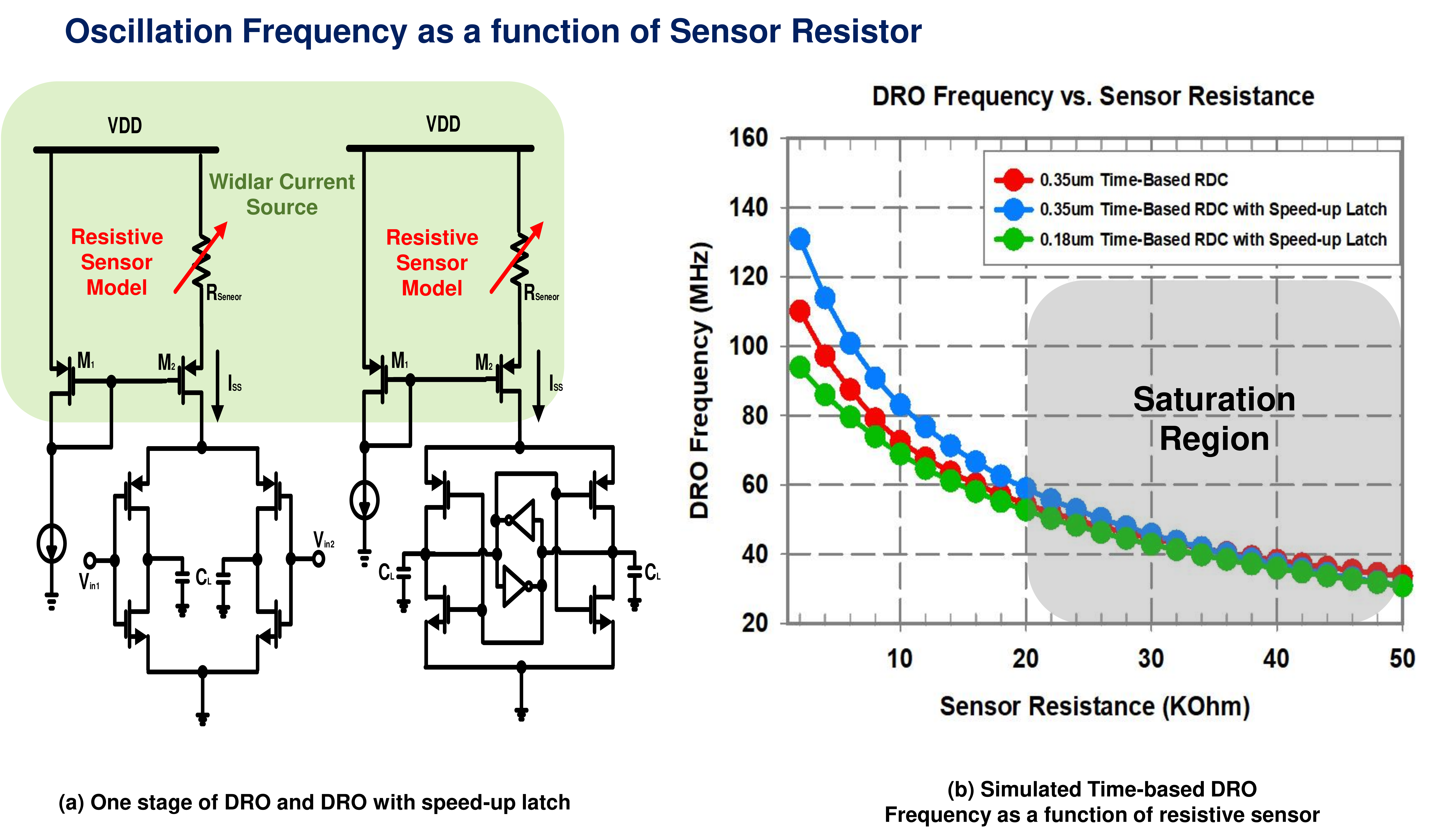}
\caption{Simulation results for the time-based RDC in 0.18$\mu$m technology: (a) one stage of the DRO and DRO with speed-up latch for analysis of \(R_{sensor}\) range and circuit design and (b) DRO frequency as a function of \(R_{sensor}\).}
\label{DROFrequency_R_Sensor}
\end{figure}
\begin{figure*}[!t]
\centering
\includegraphics[width=7in]{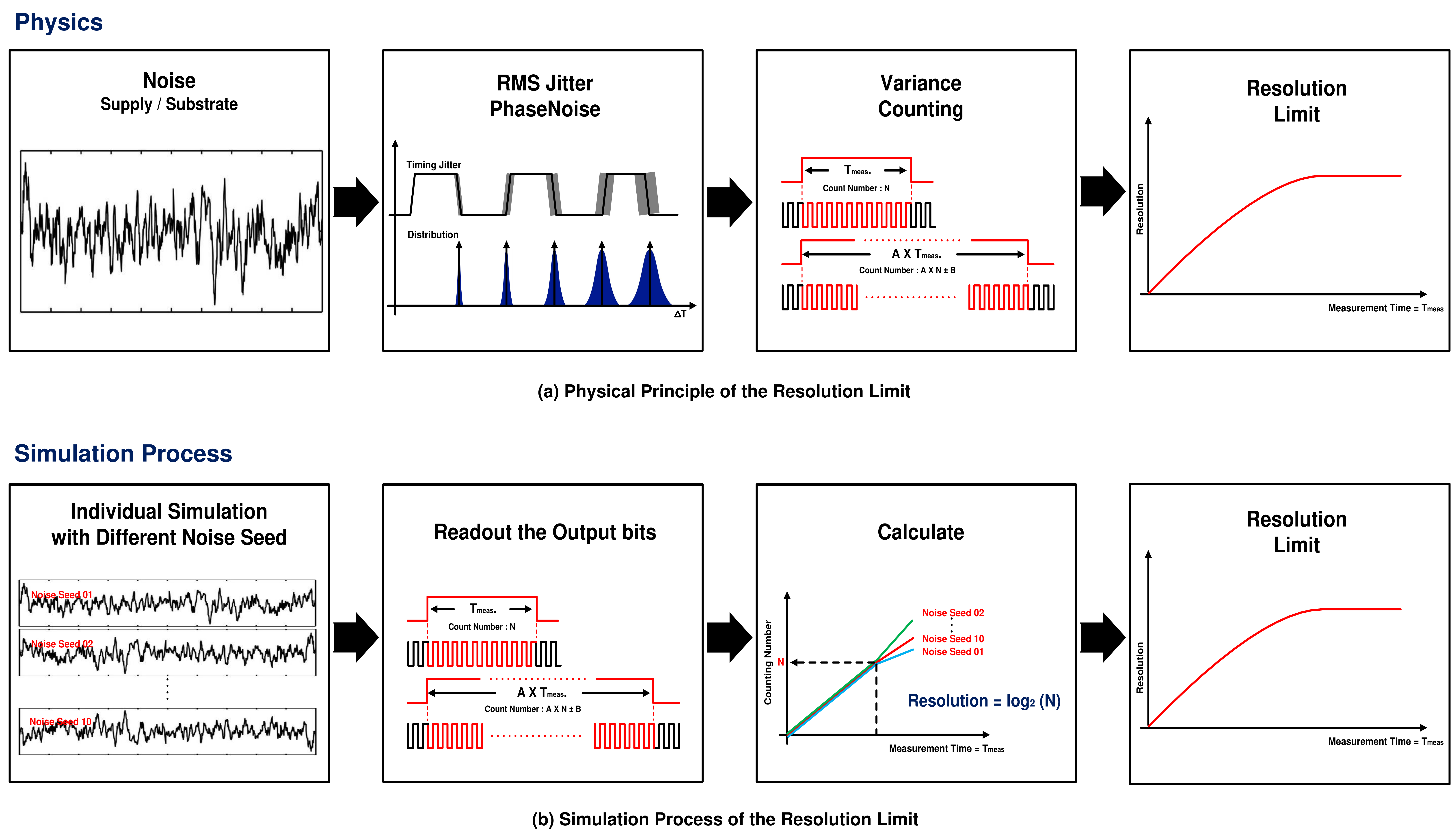}
\caption{System level simulation steps for the time-based RDC: (a) physical principle of the resolution limit and (b) simulation process to get the maximum bit-resolution.}
\label{Simulation_Step}
\end{figure*}
\subsection{Resistive Sensor of the Time-Based RDC}
A RDC measures the resistance value of a resistive sensor and the input range is a relevant parameter for a RDC. For determining the range of the sensing resistance, the Widlar current source configuration employed in the proposed design was analyzed. Fig. 7(a) shows the widlar configuration. The output current (\(I_{SS}\)) in saturation region is defined by Eq. 7 \cite{widlar}.
\begin{equation}
\begin{aligned}
	I_{SS} = \left(\frac{\sqrt{\frac{2}{\beta_{M2}}+4R_{sensor}(V_{SG,M1}-\left|V_{Tp,M1}\right |)}-\sqrt{\frac{2}{\beta _{M2}}}}{2R_{sensor}}\right)^{2}
\end{aligned}
\label{comm_eqn}
\end{equation}
\(I_{SS}\) is controlled by the resistive sensor (\(R_{sensor}\)), and oscillation frequency (which is a function of the delay of the ring oscillator stage, and hence a function of the current through the stage) becomes a function of \(R_{sensor}\). Fig. 7(b) shows the DRO oscillation frequency of the 0.35$\mu$m time-based RDC(design 1), the 0.35$\mu$m time-based RDC with speed-up latch (design 2) and the 0.18$\mu$m time-based RDC with speed-up latch (design 3) respectively, as a function of \(R_{sensor}\). The oscillation frequency of DRO is saturated beyond around 20k$\Omega$ in all three cases. Therefore, the input range of the RDCs are $<$ 20k$\Omega$

In this architecture, the resistive sensors are designed in a way that the generated frequency primarily depends on the amount of current allowed by the resistance. When the values of the three resistive sensors are the same (\(R\): fully matched scenario), the oscillation frequency would be \(f_{osc} = \frac{1}{2Nt_{p}}\), where N ia the number of stages and \(t_{p}\) is the stage delay corresponding to \(R\). On the other hand, when the values of the three resistive sensor are not the equal \((R_{1} \neq R_{2} \neq R_{3})\), the oscillation frequency would be \(f_{osc} = \frac{1}{2(t_{p1}+t_{p2}+t_{p3})}\) where \(t_{pi}\) is the \(i\)--th stage delay corresponding to \(R_{i}\). This means that the final frequency is a function of the average resistance. Due to a delta amount of change in the resistance of the sensor, DRO delay of each stage would be determined by \(R_{1}+\Delta R_{1},R_{2}+\Delta R_{2}\), and \(R_{3}+\Delta R_{3}\). The final output frequency can be expressed \(f_{osc} = \frac{1}{2\left \{(t_{p1}+ \Delta t_{p1})+(t_{p2}+ \Delta t_{p2})+(t_{p3}+ \Delta t_{p3})\right \}}\), which is again the average of each delay. Given that the resistances of all three sensors increase (or decrease) simultaneously, it is not necessary to have matching as a requirement.

\begin{figure*}[!t]
\centering
\includegraphics[width=7in]{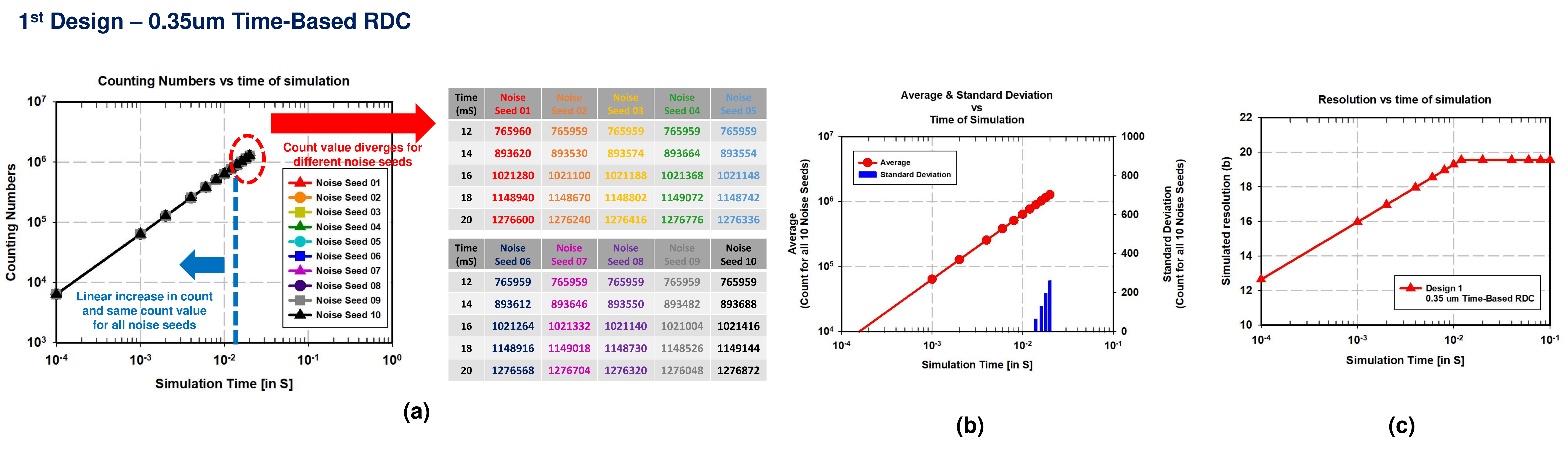}
\caption{System level Simulation results for the time-based RDC in 0.35$\mu$m technology: (a) 10 individual counting number obtained from the rising edges of output, (b) average and standard deviation of counting numbers for all 10 noise seeds and (c) resolution vs. time of simulation of the 0.35$\mu$m time-based RDC (design 1).}
\label{Simulation_Result}
\end{figure*}
\begin{figure*}[!t]
\centering
\includegraphics[width=7in]{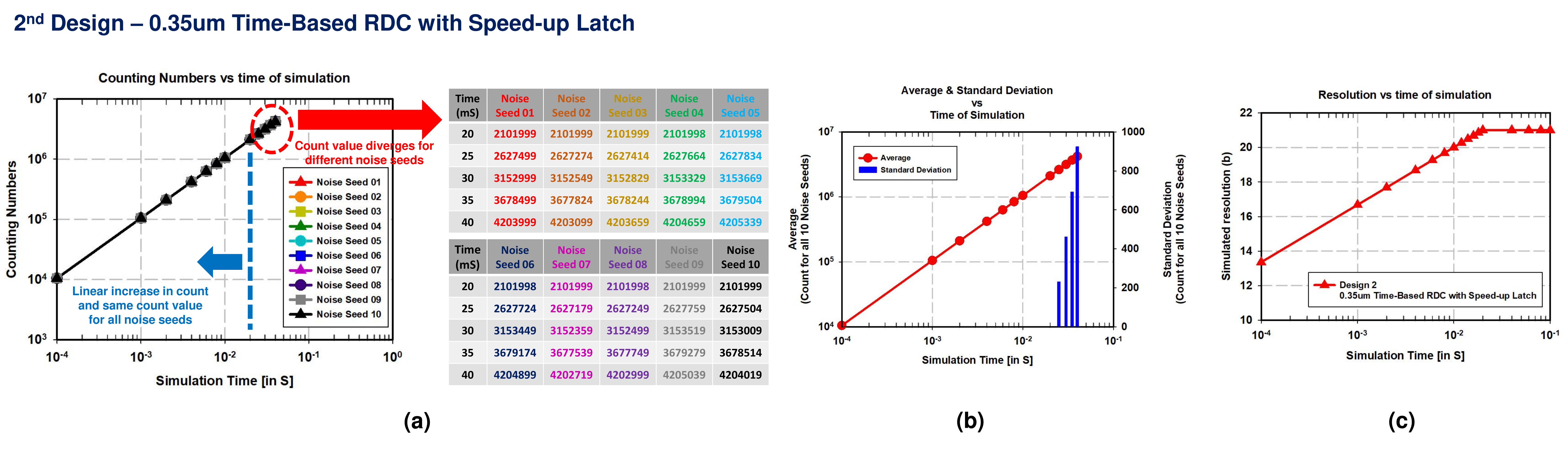}
\caption{System level Simulation results for the time-based RDC with speed-up latch in 0.35$\mu$m technology: (a) 10 individual counting number obtained from the rising edges of output, (b) average and standard deviation of counting numbers for all 10 noise seeds and (c) resolution vs. time of simulation of the 0.35$\mu$m time-based RDC with speed-up latch (design 2).}
\label{Simulation_Result}
\end{figure*}
\begin{figure*}[!t]
\centering
\includegraphics[width=7in]{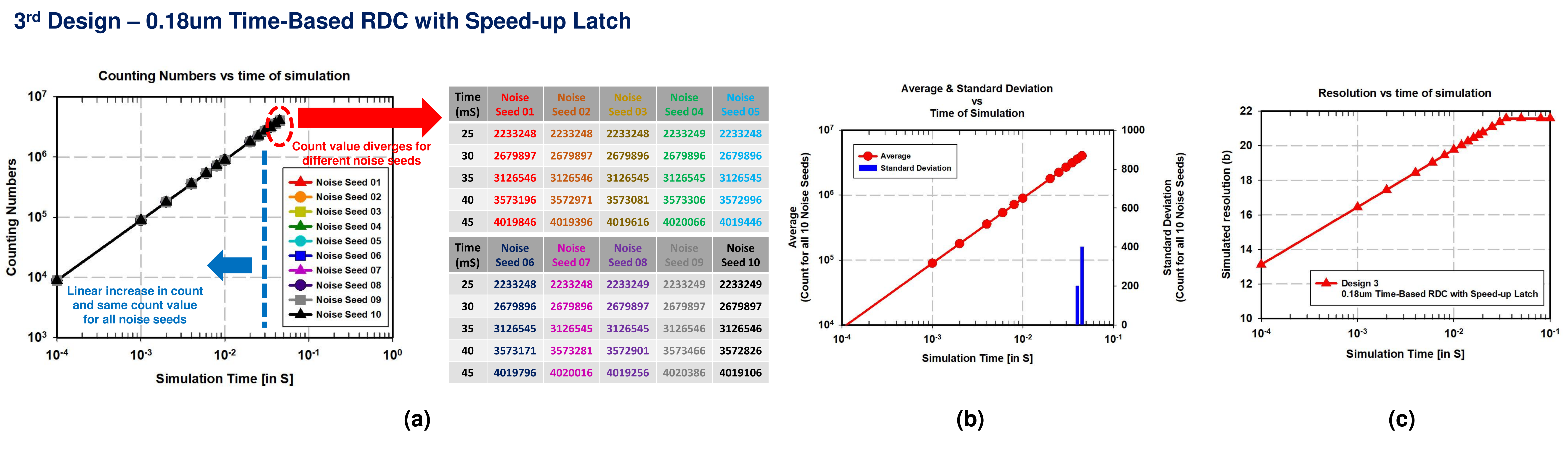}
\caption{System level Simulation results for the time-based RDC with speed-up latch in 0.18$\mu$m technology: (a) 10 individual counting number obtained from the rising edges of output, (b) average and standard deviation of counting numbers for all 10 noise seeds and (c) resolution vs. time of simulation of the 0.18$\mu$m time-based RDC with speed-up latch (design 3).}
\label{Simulation_Result}
\end{figure*}
\begin{figure}[!t]
\centering
\includegraphics[width=3.5in]{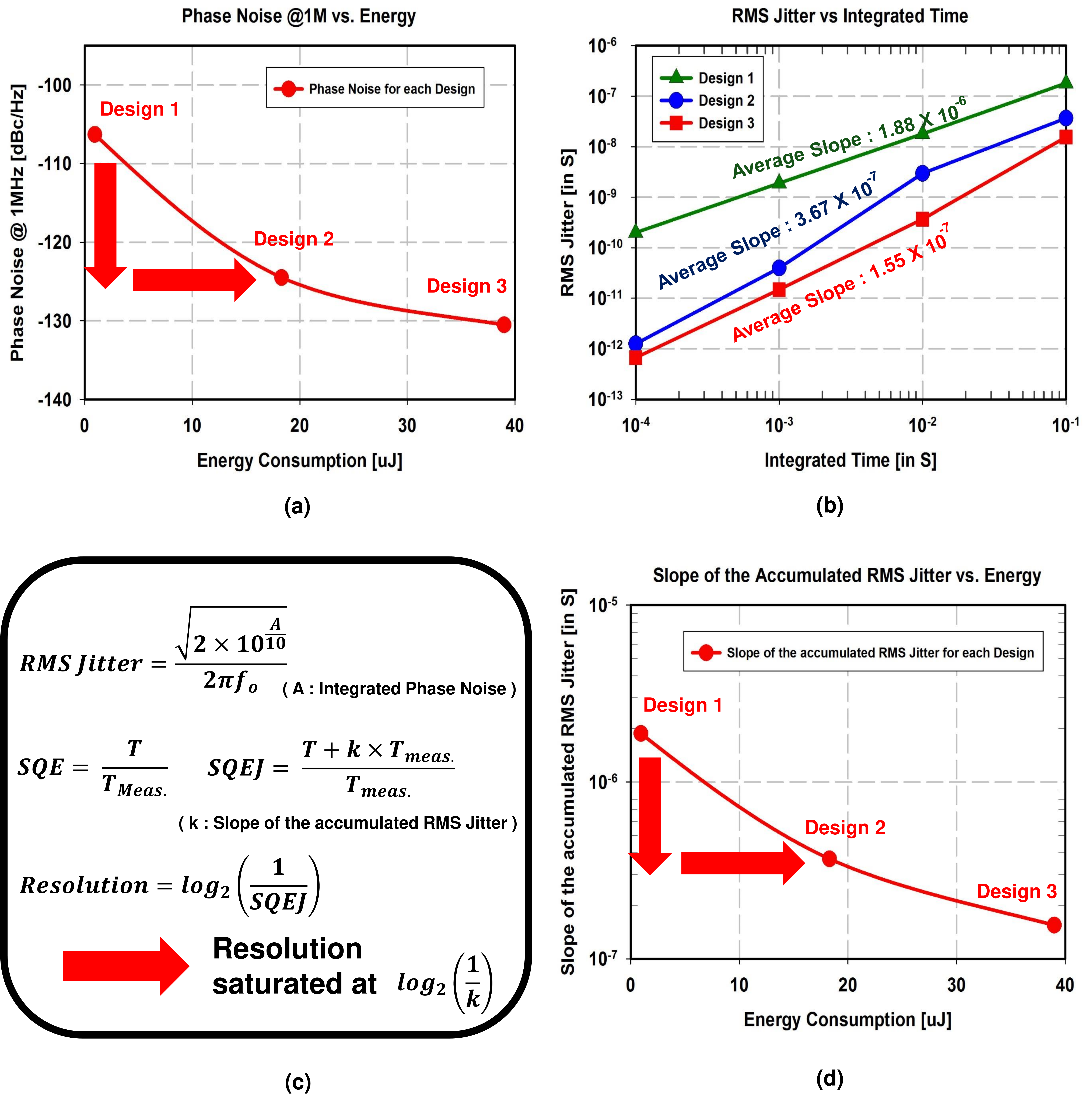}
\caption{Simulation results for the three time-based RDC: (a) phase noise vs. energy consumption, (b) the accumulated rms jitter vs. time of mreasurement, (c) equations and (d) the slope of the accumulated rms jitter vs. energy consumption.}
\label{Sim_180nm}
\end{figure}
\begin{figure}[!t]
\centering
\includegraphics[width=3.5in]{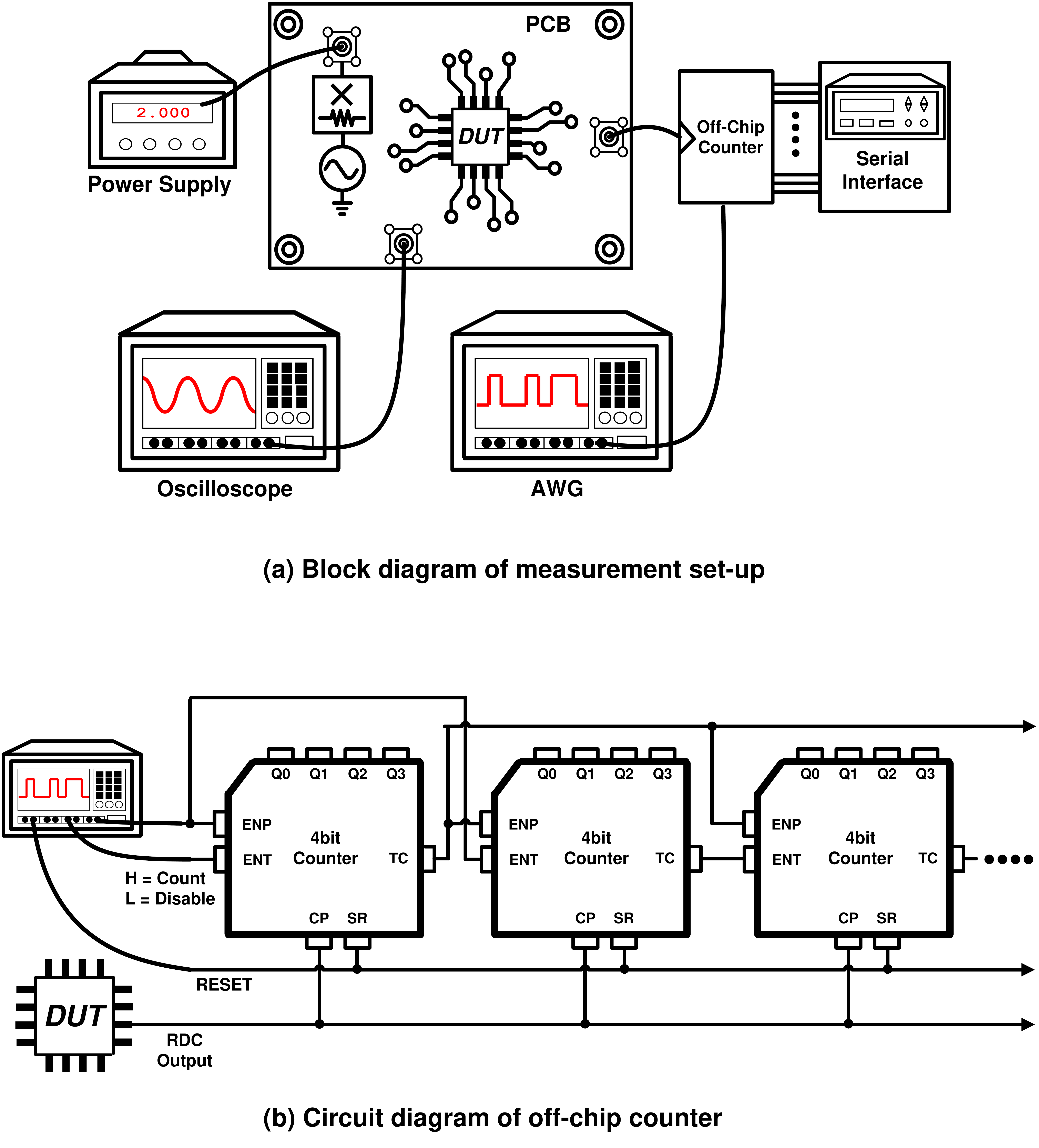}
\caption{(a) Block diagram of measurement set-up and (b) circuit diagram of off-chip counter}
\label{FigX}
\end{figure}
\begin{figure*}[!t]
\centering
\includegraphics[width=7in]{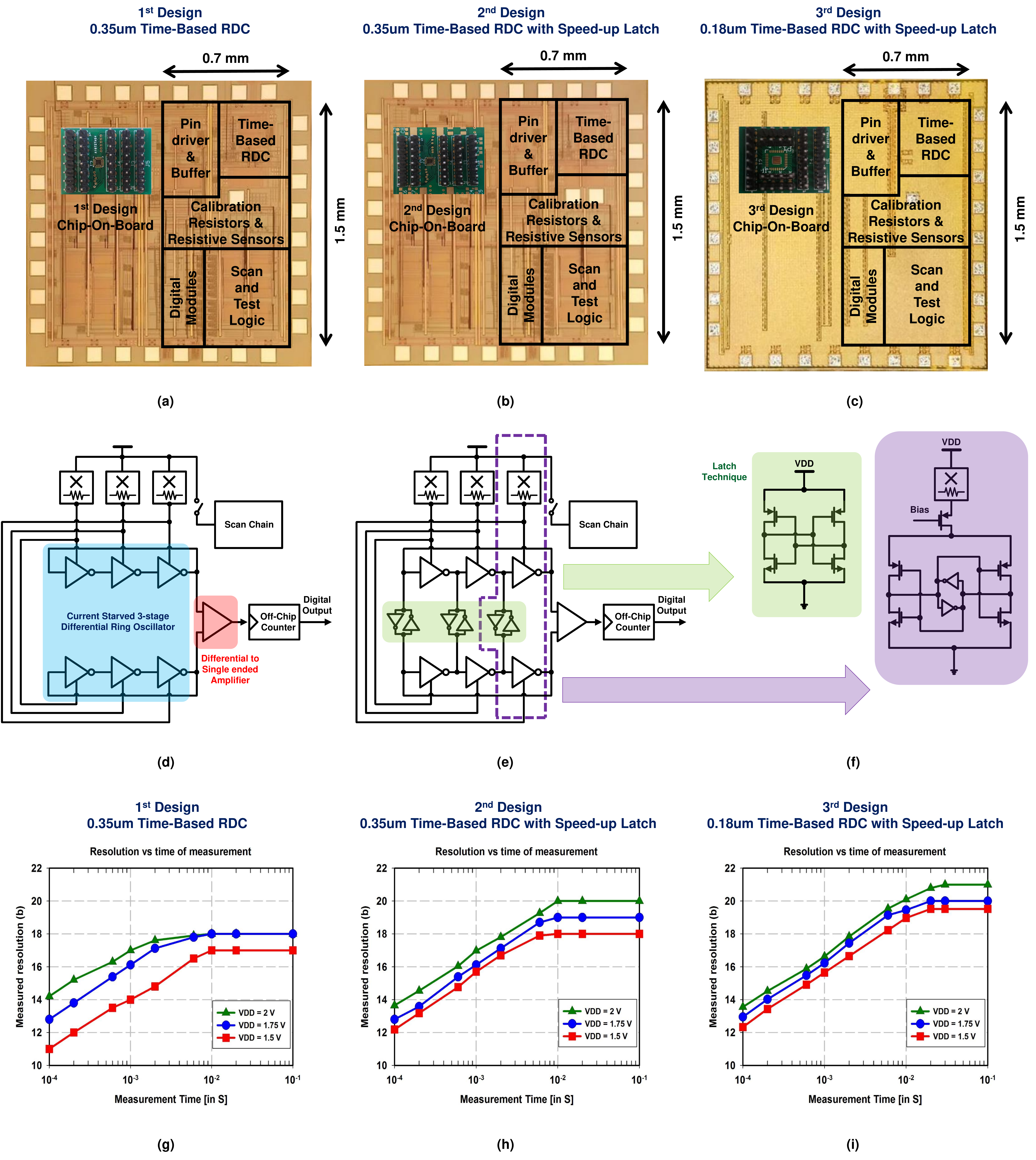}
\caption{(a) Chip micrograph of the 0.35$\mu$m time-based RDC (design 1), (b) chip micrograph of the 0.35$\mu$m time-based RDC with speed-up latch (design 2), (c) chip micrograph of the 0.18$\mu$m time-Based RDC with speed-up latch (design 3), (d) circuit diagram of 0.35$\mu$m time-based RDC, (e) circuit diagram of the 0.35$\mu$m and 0.18$\mu$m time-based RDC with speed-up latch, (f) detailed circuit schemetic consisting of the speed-up latch technique, (g) resolution vs. time of measurement of the 0.35$\mu$m time-based RDC (design 1), (h) resolution vs. time of measurement of the 0.35$\mu$m time-based RDC with speed-up latch (design 2), and (i) resolution vs. time of measurement of the 0.18$\mu$m time-based RDC with speed-up latch (design 3).}
\label{3by3_fig}
\end{figure*}
\section{System Level Simulation}
Fig. 8 describes physics of the resolution limit and the corresponding simulation process. Many noise factors influence the rms jitter of this system. For example, the rms jitter of the system is all affected by flicker noise up conversion in the tail current source and flicker noise from the correlated supply and substrate noise. The rms jitter accumulate linearly with the measurement time (\(T_{meas.}\)). The linear increase in the rms jitter was theoretically shown in \cite{jitter/phasenoise13} - \cite{jitter/phasenoise14}. However, importantly, the counting number of the rising edges of output does not linearly increase, even if the (\(T_{meas.}\)) increases, primarily due to rms jitter, which is caused by noise. This subsequently explains the resolution limit of the system, as shown in Fig. 8(a). Fig 8(b) describes the simulation process to demonstrate the limit of the bit-resolution. Simulations are computed and verified via \(Spectre^{TM}\) simulations using TSMC 0.35$\mu$m and 0.18$\mu$m technology. 10 individual transient simulation were performed with different noise seeds. The result of simulations, which refers to the rising edges of the output bits, is read at specific times. The counting numbers, represented by the 10 separate simulation results along with a different noise seed, are compared at specific times, respectively. The maximum bit-resolution can be obtained from the point where the fluctuation of the counting numbers compared becomes severe.

Fig. 9, Fig. 10 and Fig. 11 each shows the system level simulation result of the 0.35$\mu$m time-based RDC(design 1), the 0.35$\mu$m time-based RDC with speed-up latch (design 2) and the 0.18$\mu$m time-based RDC with speed-up latch (design 3) respectively. Fig. 9(a), Fig. 10(a) and Fig. 11(a) show that the counting number, which can be obtained from the rising edges of output, increases with the simulation time. The counting number linearly increases and same count value for all noise seeds with the simulation time until it reaches at a certain simulation time, 12ms, 20ms and 35ms, respectively. However, beyond that simulation time, the count value diverges for different noise seeds; rather, it changes randomly in all three cases. This happens for two reasons. One is that 10 different noise seeds were implemented to each individual simulation. The other is that the oscillation frequency of this system constantly changes, either accelerating or deceleration, due to noise. In order to show the result of simulation, both the average and standard deviation of the counting number, which derived from 10 individual transient simulation performed with different noise seeds, were calculated as shown in Fig 9(b), Fig 10(b) and Fig. 11(b). Up until the specific simulation time of 12ms, 20ms and 35ms, the standard deviation for the counting number is zero. However, beyond that simulation time, the standard deviation increases as the simulation time increases in all three cases. The bit-resolution of the three designs of time-based RDC is plotted against the time of simulation in Fig. 9(c), Fig. 10(c) and Fig. 11(c), respectively. The results show a linear increase in resolution with simulation time on log scale until they saturate in all three cases. The maximum bit resolution can be calculated using Eq.8
\begin{equation}
\begin{aligned}
	bit\textunderscore resolution = log_{2}(N)
\end{aligned}
\label{comm_eqn}
\end{equation}
where N refers to the maximum counting number of the rising edges of output before it significantly changes. This simulation show that the maximum counting number of the rising edges of output for the design1, the design 2 and the design 3 is 765,959, 2,101,999 and 3,126,546 respectively. As a result, the maximum bit-resolution of the design1 and the design 2 and the design3 can be achieved 19.54 bits with 12ms, 21 bits with 20ms, and 21.55 bits with 35ms, respectively. To check the sensitivity of the DRC with respect to voltage and temperature, 10 individual transient simulation were performed with different noise seeds and different supply voltage and temperature. Supply voltage is changing 1.5V-2V and temperature is changing -25$^\circ$C to 100$^\circ$C. The worst case variations with in 17.2 to 19.3 bits in design1, 19.5 to 21.4 bits in design2, and 20.2 to 21.9 bits in design3.

Fig. 12(a) shows the simulation results for the relationship between phase noise and energy consumption of the three designs. The phase noise from design 1 to design 3 is improved even though, energy consumption increased significantly. Based on this phase noise simulation results, the accumulated RMS jitter for integrated time can be calculated. The RMS jitter is calculated using Eq.9
\begin{equation}
\begin{aligned}
	RMS Jitter = \frac{\sqrt{2\cdot 10^{A/10}}}{2\cdot \pi\cdot f_{o}}
\end{aligned}
\label{comm_eqn}
\end{equation}
where A refers to the integrated phase noise power, and \(f_{o}\) is the oscillation frequency. Fig. 12(b) plots the accumulated RMS jitter over the integrated time. The maximum achievable resolution can be calculated from the slope of the accumulated RMS jitter. The value of slope for the design 1, the design 2 and the design 3 is 1.88 X \(10^{-6}\), 3.67 X \(10^{-7}\) and 1.55 X \(10^{-7}\) respectively. For calculating the average slope, two points at 0.1ms and 10ms are considered. Even though integrated time increases, the bit-resolution is eventually saturated at \(log_{2}(\frac{1}{k}) \) as shown in Fig. 12(c). Fig. 12(d) presents the relationship between the slope of the accumulated RMS jitter and energy consumption of the three designs. Although the energy consumption is increasing gradually, the decrease in the slope of the accumulated RMS jitter dose not keep up with it and is gradually saturated. Our application, we have put priority on resolution even at the expense of energy consumption and hence we are exploring how high can we go in resolution. From the system simulation results, The 0.35$\mu$m time-based RDC targeted towards maximizing the energy/conversion step, while the 0.18$\mu$m time-based RDC with speed-up latch targeted the highest resolution.

\section{System Level Measurement Results}
Fig. 13(a) describes the block diagram of measurement setup with 3 designs of energy-resolution scalable time-based RDC. The implemented time-based RDC chip was measured by using chip-on-board (COB) setup on a customized Printed Circuit Board (PCB) with wire-bonding. The output of the time-based RDC is connected with an off-chip counter. The off-chip counter is implemented with 6 synchronous 4bit binary counters, which are connected in cascade. The synchronous 4 bit binary counter is manufactured by Texas Instruments, with the device model is named as "CYFCT163T". The "CYFCT163T" device has two types of count enable (CET and CEP) inputs, a terminal count(TC) output for versatility in forming synchronous multistaged counter, and synchronous reset(SR) input. Fig. 13(b) shows the circuit diagram of off-chip counter. This diagram demonstrates how the counter can be used to implement a high-speed N-bit counter. To avoid reducing maximum clocking rate by adding additional stages due to the propagation delay of the terminal out, the output-look-ahead circuit structure is implemented. When additional stages are added the maximum clock frequency of the counter, \(CLK_{MAX}\) remains unchanged. The \(CLK_{MAX}\) is defined by Eq.10.
\begin{equation}
\begin{aligned}
	CLK_{MAX} = \frac{1}{TCt_{PHL}} + CEPt_{su}
\end{aligned}
\label{comm_eqn}
\end{equation}
where \(t_{PLH}\) is the propagation delay time, low to high level output, and \(t_{su}\) is the setup time. The output bit of each 4-bit binary counter is connected with serial interface (National Instrument). The generated frequency from the RDC provides the clock signal of the off-chip counter. During measurement time \(T_{meas.}\), the counter counts the rising edges of output. The counter output represents the integer number of output cycles during one readout within the predefined measurement time \(T_{meas.}\). The measurement time signal is generated from a RIGOL DG4200 Arbitrary Waveform Generator (AWG). Alternatively, a serial interface with a microcontroller could be utilized for the readout.

Fig. 14(a), (b) and (c) show micro-photograph and Chip-on-board (COB) of the implemented 0.35$\mu$m time-based RDC (design 1), 0.35$\mu$m time-based RDC with speed-up latch (design 2), and the 0.18$\mu$m time-based RDC with speed-up latch (design 3). The active area of the chip of all 3 design is less than 1.1 mm\(^{2}\) excluding pads. Fig. 14(d) and (e) present the implemented circuit diagram of each time-based RDC. Fig. 14(f) describes the detailed circuit schemetic that consists of the speed-up latch technique. The speed-up latch improve the slope of rising and falling edge of a 3-stage ring oscillator and as a result, enhances the phase noise performance by improving the swing. The bit-resolution of the three designs of time-based RDC is plotted against the time of measurement in Fig 14(g), (h) and (i), respectively. The results show a linear increase in resolution with measurement time on log scale for 3 supply voltages until they saturate as a result of jitter/phase noise accumulation, as explained in \cite{CICC7}-\cite{JSSC7}. The 0.35$\mu$m time-based RDC with speed-up latch (design 2) increases the slope of the rising and falling edges by providing a positive feedback of the output of the latch. Compared to 0.35$\mu$m time-based RDC (design 1), the phase noise is improved which subsequently results in higher bit-resolution. The 0.18$\mu$m time-based RDC with speed-up latch (design 3) reduces the power consumption for similar readout time. As compared to design 2, 1-bit better resolution can be achieved when readout time is increased to 30ms. The three designs of energy-resolution scalable time-based RDC achieve 18 bit-resolution at 861nW \cite{CICC7}-\cite{JSSC7}, 20 bit-resolution at 19.1$\mu$W  and 21 bit-resolution at 52.8$\mu$W , respectively (design 1-2 with 10ms readout time, and design 3 with 30ms readout time, with one readout every second).
\begin{figure}[!t]
\centering
\includegraphics[width=3.5in]{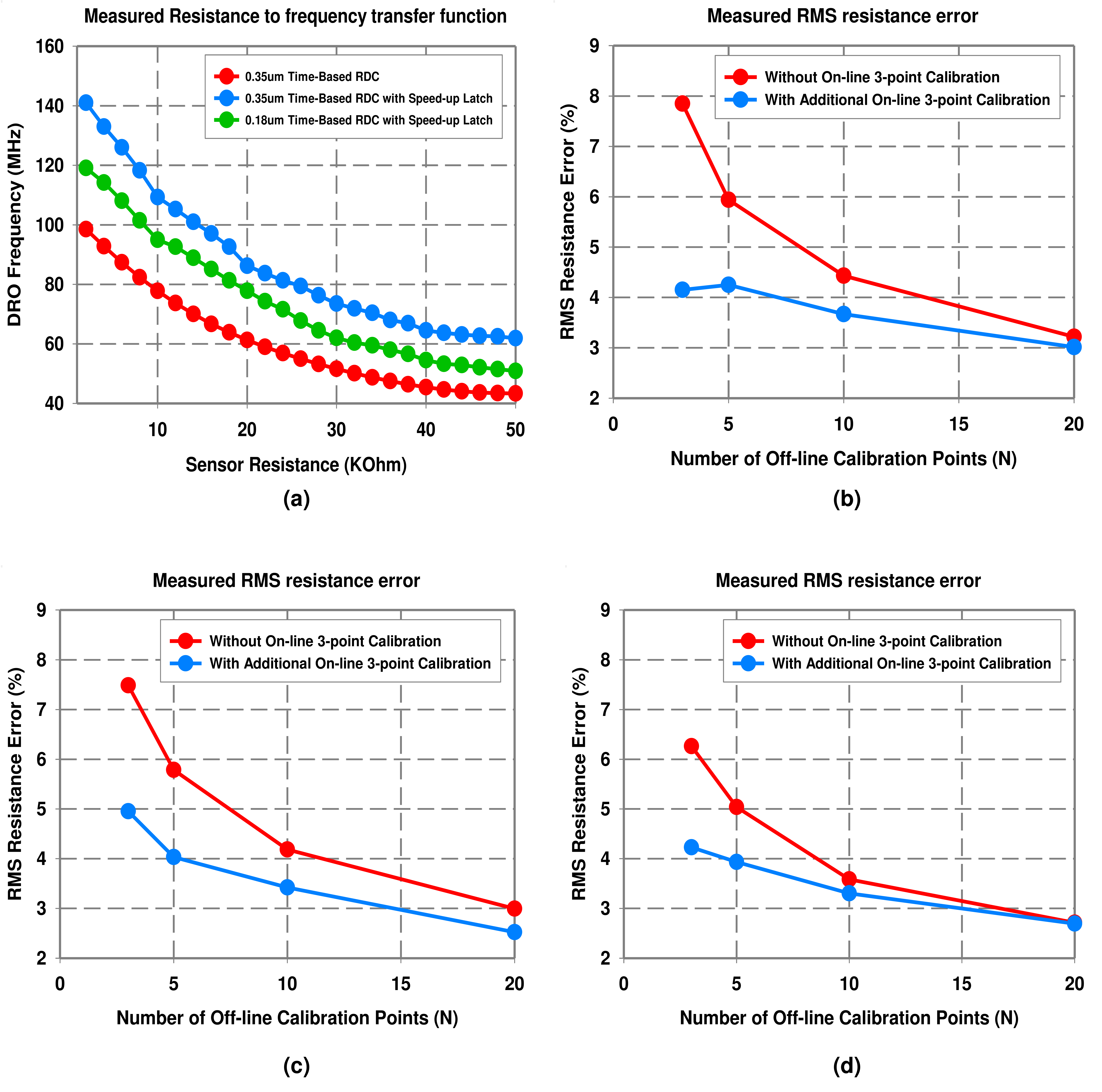}
\caption{(a) Measured resistance to frequency transfer function of the 3 RDC designs, (b) measured RMS resistance error (\%) for Per-device Off-line calibration, followed by On-line 3-point calibration for the 0.35$\mu$m time-based RDC(design 1), (c) measured RMS resistance error (\%) for Per-device Off-line calibration, followed by On-line 3-point calibration for the 0.35$\mu$m time-based RDC with speed-up latch(design 2) and (d) measured RMS resistance error (\%) for Per-device Off-line calibration, followed by On-line 3-point calibration for the 0.18$\mu$m time-based RDC with speed-up latch(design 3)}
\label{calibration}
\end{figure}
\begin{table*}[!t]
\caption{Measured performance summary of the 3-design time-based RDC and comparison table}
\label{table_example}
\centering
\begin{tabular}{|
>{\columncolor[HTML]{EFEFEF}}c |
>{\columncolor[HTML]{ECF4FF}}c |
>{\columncolor[HTML]{ECF4FF}}c |
>{\columncolor[HTML]{ECF4FF}}c |
>{\columncolor[HTML]{FFCCC9}}c |
>{\columncolor[HTML]{FFCCC9}}c |
>{\columncolor[HTML]{FFCCC9}}c |
>{\columncolor[HTML]{FFCCC9}}c |
>{\columncolor[HTML]{FFCCC9}}c |}
\hline
\cellcolor[HTML]{9B9B9B}Parameter & \multicolumn{3}{c|}{\cellcolor[HTML]{34CDF9}This Work : 3-design RDC} & \cellcolor[HTML]{FD6864}\cite{table1} & \cellcolor[HTML]{FD6864}\cite{table2} & \cellcolor[HTML]{FD6864}\cite{table3} & \cellcolor[HTML]{FD6864}\cite{table4} & \cellcolor[HTML]{FD6864}\cite{table5} \\
\hline &
\begin{tabular}[c]{@{}c@{}}Time-Based\\ RDC\end{tabular} & \begin{tabular}[c]{@{}c@{}}Time-Based\\ RDC with\\ speed-up latch\end{tabular} & \begin{tabular}[c]{@{}c@{}}Time-Based\\ RDC with\\ speed-up latch\end{tabular} & \begin{tabular}[c]{@{}c@{}}Time-Based\\ ADC\end{tabular} & \begin{tabular}[c]{@{}c@{}}Time-Based\\ ADC\end{tabular} & \begin{tabular}[c]{@{}c@{}}SB-PM\\ RCDC\end{tabular} & RDC & RDC\\
\hline
Technology & 0.35$\mu$m & 0.35$\mu$m & 0.18$\mu$m & 0.35$\mu$m & 40 nm & 0.18$\mu$m & 0.18$\mu$m & 65 nm\\
\hline
Architecture & \multicolumn{3}{c|}{\cellcolor[HTML]{ECF4FF}Current Starved Ring Oscillator Based} & VCO Based & VCO Based &
\begin{tabular}[c]{@{}c@{}}Oscillator\\ Based\end{tabular} & SAR & \begin{tabular}[c]{@{}c@{}}Switched\\ Capacitor\end{tabular} \\
\hline
Supply Voltage & 1.75 V & 2 V & 2 V & 5V, 1.8 V & 1.2 V, 0.45 V & 1 V & 1.8 V & 1 V \\
\hline
Power & 86.1 uW & 1.92 mW & 1.76 mW & 340 uW & 7 uW & 140 uW & 93.2 uW & 12.3 uW \\
\hline
Resolution & 18 bit & 20 bit & 21 bit & 12 bit & 12 bit & 16.6 bit & 11.3 bit & 9 bit \\
\hline
Readout Time & 10 ms & 10 ms & 30 ms & 400 ms & 0.75 ms & 2.93 ms & 0.92 ms & 0.5 ms \\
\hline
Dynamic Range & 103.7 dB & 113.5 dB & 121.1 dB & 73 dB & 79 dB & N/A & N/A & N/A \\
\hline
\begin{tabular}[c]{@{}c@{}}Off-chip Counter\\
Power\end{tabular} & N/A & 240 mW & 205 mW & N/A & N/A & N/A & N/A & N/A \\
\hline
\begin{tabular}[c]{@{}c@{}}On-chip Counter\\
Power\end{tabular}  & 14 uW  & 14.3 uW & 10.1 uW & N/A & N/A & N/A & N/A & N/A \\
\hline
Energy & 861 nJ & 19.2 uJ & 52.8 uJ & 136 uJ & 5.263 nJ & 410 nJ & 87.744 nJ & 6.15 nJ \\
\hline
FoM(Energy/cs) & 3.29 pJ/cs & 18.3 pJ/cs & 25.1 pJ/cs & 37250 pJ/cs & 4.27 pJ/cs & 4.04 pJ/cs & 33 pJ/cs & 12 pJ/cs \\
\hline
Chip Area & \multicolumn{3}{c|}{\cellcolor[HTML]{ECF4FF}0.435 mm\(^{2}\) (RDC), 1.05mm\(^{2}\) (total)} & 0.36 mm\(^{2}\) & 0.135 mm\(^{2}\) & 0.175 mm\(^{2}\) & 0.27 mm\(^{2}\) & N/A \\
\hline
\end{tabular}
\end{table*}

The fundamental nature of the resistance to frequency transfer function is non-linear as shown in \cite{JSSC7}, and the measured result for this transfer function for the 3 designs is shown in Fig. 15(a). This non-linearity arises from the ring-oscillator stage in the Widlar current source configuration which is utilized to convert the degeneration resistance to a corresponding delay and  hence to the frequency of the ring oscillator. Because the input signal is of very low frequency (or effectively almost a DC quantity as shown in the application of [8]), the dynamic range can be found from the range of the frequencies (or range of count values from the RDC). This range is found to be about 103dB for the original 0.35$\mu$m time-based RDC(design1) \cite{JSSC7}, 113.5dB for the 0.35$\mu$m time-based RDC with the speed-up latch(design2), and 121.1dB for the 0.18$\mu$m time-based RDC with the speed-up latch(design3). However, because of the non-linear nature of the resistance to frequency transfer function, calibration and post-processing is required for the proposed RDC. Error due to this calibration becomes a dominant factor in the performance and linearity representation. Hence, instead of performing a traditional FFT (or DNL/INL analysis) to represent linearity, we analyze the effect of the non-linearity calibration method on the overall readout error.The calibration/correction for the non-linearity can be done in 3 ways:
\begin{enumerate}[1)]
	\item \emph{Per-device Off-line calibration:}\\ This requires finding out the resistance to frequency transfer function for each device during a pre-measurement (off-line) calibration and applying the inverse of that function during measurement to correct for the non-linearity. This method would result in better accuracy as the number of points used for calibration increase. In the limiting case, error will tend to zero (resulting in extremely high linearity) as the number of points approach infinity. However, this incurs high amount of cost in terms of time and available manual resources.
	\item \emph{2)	Per-batch Off-line calibration, followed by an On-line calibration:}\\ This requires finding out the resistance to frequency transfer function off-line, for one device out of a batch of devices, and eventually update the transfer function for each device on-line, during measurement to take care of PVT (process, voltage and temperature) variations. However, results might be largely inaccurate due to small number of on-line data points for calibration and large process variations.
	\item \emph{3)	Subset of Per-device Off-line calibration, followed by an On-line calibration:}\\ As a compromise between methods 1 and 2, we can perform per-device off-line calibration with a reduced number of points (that will lower the test cost) and then perform an on-line update of the transfer function during measurement to take care of the VT (voltage and temperature) variations.
\end{enumerate}
Fig. 15(b), (c) and (d) shows the measured RMS resistance error (\%) for Per-device Off-line calibration, followed by On-line 3-point calibration for 3 designs. For calculating the RMS resistance error, 5 frequency points were randomly selected from all frequencies which correspond to resistances within the 2kohm-50kohm  range. The inverse of the resistance to frequency transfer function was applied on these frequencies to find out the resistance. This resistance is compared with the original resistance values to find the RMS error (\%) for the 5 points. During this process, the points used for calibration and for test was always kept separate. The number of off-line calibration points are varied from 3 to 20, while the number of on-line calibration points is set to 3 (which is a feature of all 3 designs, and is shown in detail in [8]). With the additional automatic on-line calibration, the number of points required for off-line calibration reduces for similar amount of rms error.

\begin{figure}[!t]
\centering
\includegraphics[width=3.5in]{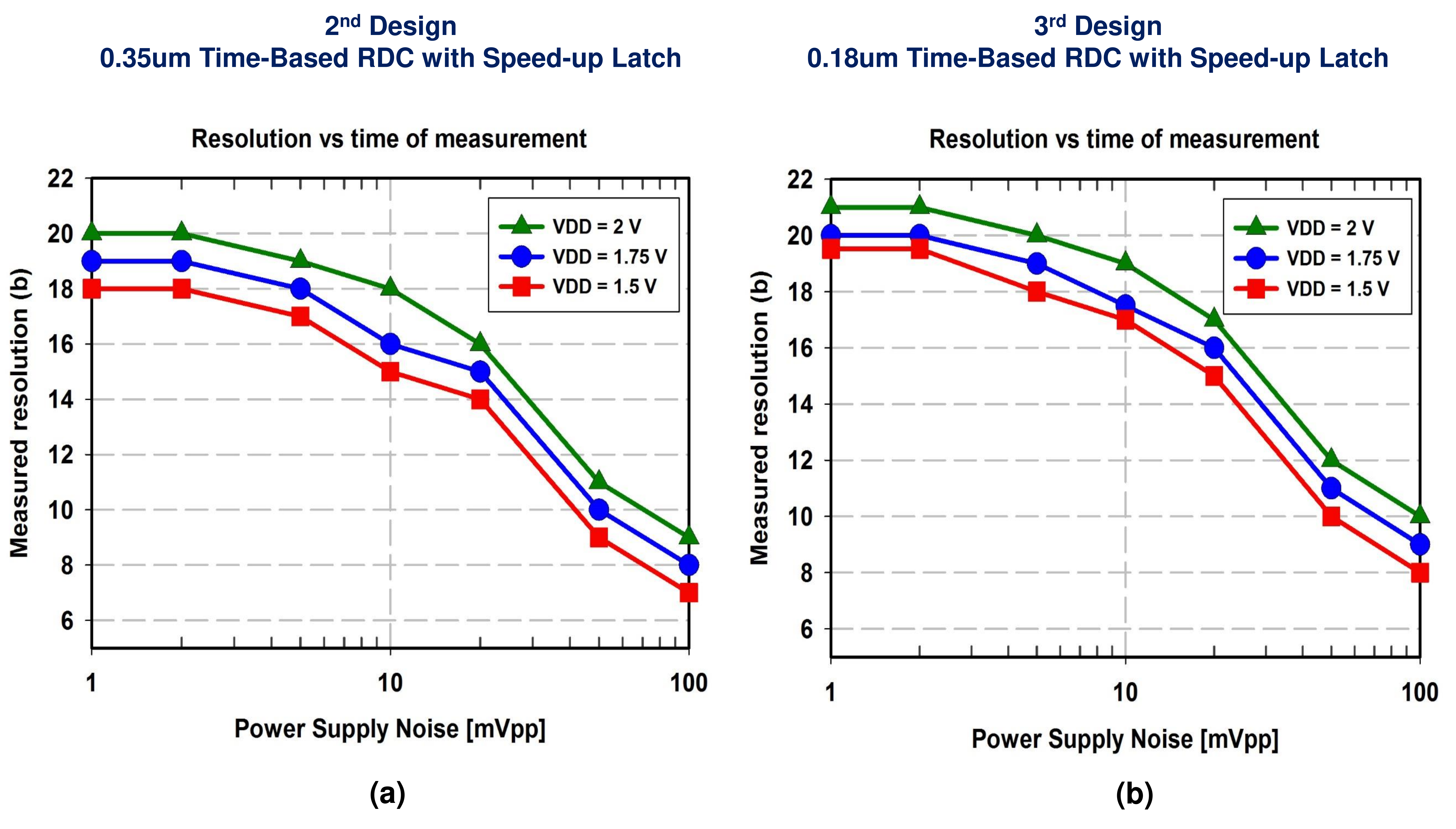}
\caption{Resolution vs. supply noise measurement (a) 0.35$\mu$m time-based RDC with speed-up latch (design 2), and (b) 0.18$\mu$m time-Based RDC with speed-up latch (design 3), showing $>18$ bit resolution is possible even with 10mV (peak-to-peak) supply noise.}
\label{noise_measurement}
\end{figure}

Table I summarizes the performance of the three designs of energy-resolution scalable time-based RDC in comparison with state-of-the-art high resolution time-based ADC architectures. The 0.35$\mu$m time-based RDC consumes the lowest energy, which is 861 nJ with 10 mSec, among all ADCs and the 0.18$\mu$m time-based RDC with speed-up latch offers the highest resolution, which is 21 bit with 30 mSec, among all ADCs. From the perspective of the energy/conversion step, 0.35$\mu$m time-based RDC shows the best performance which is 3.29 pJ/bit, among all ADCs. The existing time-based approaches utilize over-sampling the input signal for better resolution. However, the proposed approach converts input signal to a corresponding frequency and then measure this frequency over a longer period of time, thereby under-sample it. One primary advantage of under sampling is low power. Additionally, the proposed time based RDC (design 1) has the advantage of energy-resolution scalability -- a higher resolution can be obtained simply by measuring the DRO frequency over longer time, while measurement over a shorter time results in a lower resolution, but with a lower amount of of energy. Traditional time-based ADCs would need to change the frequency of the system clock to achieve energy-resolution scalability -- thereby adding additional complexity to the system. In conclusion, the basic time-based RDC has the best energy/conversion step and the time-based RDC with speed-up latch improves rms jitter/phase noise to achieve high bit-resolution.

Among the factors that degrade the phase noise performance of oscillators, power supply noise is one of the most dominant in terms of its effect on both the frequency and phase of the oscillator. The power supply voltage affects delay of ring oscillator. In a current starved ring oscillator, power supply noise will reflect as current fluctuations. Fig. 15 shows the bit-resolution of the time-based RDC with respect to the power supply noise amplitude. As the amplitude of the power supply increases, the bit-resolution decreases for all the designs. However, even with 10mV peak-to-peak supply noise, the resolution of the second and third design of RDCs remain $>18$-bit, which was the phase-noise limit of design 1. When the sensors are powered from a battery for wearable applications, the effects of supply noise would be much lower. 
\section{Conclusion}
\label{conclusion}
In this paper, we presented the design and analysis of a wearable CMOS biosensor with 3 designs of energy-resolution scalable time-based resistance to digital converter (RDC). The implemented RDC consisted of a current starved differential ring oscillator, a differential to single ended amplifier, and a off-chip counter in order to convert the change in resistance to equivalent frequency. To the best of the authors' knowledge, the 0.35$\mu$m time-based RDC is the lowest-power time-based ADC reported till date, while the 0.18$\mu$m time-based RDC with speed-up latch offers the highest resolution. Insights on the energy-resolution trade-offs, scalability aspects and effects of power supply noise are also discussed in the paper. For the low frequency signal, this implemented RDC can detect with high resolution. We wanted to explore the power/performance trade-off in experiment through 3 different design variations, tapeout and IC measurements. As future work, architectures to improve the scaled quantization error would be explored, which would achieve similar resolution with a lower readout time, thereby improving the energy/conversion step.

\end{document}